\documentclass[11pt]{article}
\usepackage{psfrag}
\usepackage{graphicx}
\usepackage{graphics}
\newcommand{\as}{a\!\!\!/}
\newcommand{\ks}{k\!\!\!/}

\newcommand{\ol}{\overline}

\date{}
\begin{document}
\baselineskip=18.6pt plus 0.2pt minus 0.1pt \makeatletter
%\@addtoreset{equation}{section} \renewcommand{\theequation}{\thesection.%
%\arabic{equation}} \makeatletter \@addtoreset{equation}{section}

%\begin{titlepage}

\title{\vspace{-3cm}
\hfill\parbox{4cm}{\normalsize \emph{LPHEA 02-03}}\\
 \vspace{1cm}
{\bf Relativistic electronic dressing in laser-assisted
electron-hydrogen collisions. I . Elastic collisions.}}
 \vspace{2cm}

\author{Y. Attaourti \thanks{e-mail: attaourti@ucam.ac.ma},
 B. Manaut \thanks{e-mail: bmanaut@phea.ucam.ac.ma}\\
 {\it {\small Laboratoire de Physique des Hautes Energies et
d'Astrophysique,}}  \\{\it {\small Facult\'e des Science Semlalia,
Universit\'e Cadi Ayyad, Marrakech, BP : 2390, Maroc. }}}
\maketitle \setcounter{page}{1}
%\end{titlepage}
\begin{abstract}
We study the effects of the relativistic electronic dressing in
laser-assisted electron-hydrogen atom elastic collisions. We begin
by considering the case when no radiation is present. This is
necessary in order to check the consistency of our calculations
and we then carry out the calculations using the relativistic
Dirac-Volkov states. It turns out that a simple formal analogy
links the analytical expressions of the differential cross
section without laser and the differential
cross section in presence of a laser field\\
\vspace{.04cm}\\
 PACS number(s): 34.80.Qb, 12.20.Ds
\end{abstract}
%\end{titlepage}

%************************************************************************
\section{Introduction}
%************************************************************************
\hspace{0.6cm} Recently, the study of relativistic aspects of
laser-induced processes has proved necessary particularly as a
result of very paramount breakthroughs in laser technology which
is capable now to attain considerable ultrahigh intensities which
could never have been dreamt of three or four decades ago. Many
experiments that have shown a relativistic signature have been
recently reported. To name some, the transition between Thomson
and Compton scattering inside a very strong laser field was
investigated by C.I. Moore, J.P. Knauer and D.D. Meyerhofer
\cite{1}. C. Bula et al. \cite{2} performed experiments on non
linear Compton scattering at SLAC. Also, there are many other
types of laser-assisted processes in which relativistic effects
may be important. For instance, the process of emission of very
energetic electrons and ions from atomic clusters which are
submitted to ultrastrong infrared laser pulses \cite{3}. It is
now obvious that the whole apparatus and formalism of the non
relativistic quantum collision theory \cite{4} has to be
revisited in order to extend known non relativistic results to
the relativistic domain. Many theoretical studies of
laser-assisted electron-atom collision have been mainly carried
out in the non relativistic regime. \cite{5}.\\ \indent In
presenting this work, we want to show that the modifications of
the relativistic differential cross section corresponding to the
elastic collision $e^-+H(Is_{1/2})\rightarrow e^-+H(Is_{1/2})$
due to the dressing of the Dirac-Volkov electron, in the presence
of an ultraintense laser field can provide many interesting
insights concerning the importance and the signatures of the
relativistic effects. \\ \indent In this work, we do not consider
very high laser intensities that allow pair creation \cite{6}and
focus instead on the domain of intensities that justifies a strong
classical electromagnetic potential \cite{7}. The Dirac-Volkov
electrons are thus dressed by a strong classical electromagnetic
field with circular polarization. \\
 \indent The organization of
this paper is as follows :\\
\indent In section (2) we will present the formalism and
establish the expression of the relativistic differential
scattering cross section in absence of the laser field. This will
serve as a guide and test the consistency of the calculation in
presence of the laser field. In section (3), we give the
expression of the differential cross section in presence of a
laser field with circular polarization and we compare it with the
relativistic differential cross section without laser field. We
show that a simple formal
analogy links these two differential cross sections.\\
 \indent In section (4), we give a brief discussion of the
results. In section (5), we give a brief conclusion. Throughout
this work, we use atomic units and the abbreviation DCS stands
for differential cross section.
%\end{titlepage}
%***********************************************************************
\section{Differential cross section without laser field}
%************************************************************************
In order to recover the relativistic differential cross section
without the laser field, we begin by considering the process
$e^-+H(Is_{1/2})\rightarrow e^-+H(Is_{1/2})$ in the absence of
radiation. The (direct) transition amplitude corresponding to
this process is :\begin{eqnarray}
S_{fi}&=&-i\int_{-\infty}^{+\infty}dt<\psi_{p_f}({\mathbf r}_1,t)\phi_f({\mathbf r}_2,t)|V_d|\psi_{b_i}({\mathbf r}_1,t)\phi_i({\mathbf r}_2,t)>\nonumber\\
&=&-\frac{i}{c}\int d^4x_1\ol{\psi}_{p_f}({\mathbf
r}_1,t)\gamma^0\psi_{p_i}({\mathbf r}_1,t)<\phi_f({\mathbf
r}_2)|V_d|\phi_i({\mathbf r}_2)>\label{1}
\end{eqnarray}
Where $\psi_p({\mathbf r},t)=u(p,s)e^{-ip.x}/\sqrt{2EV}$ is the
electron wave functions described by a free Dirac spinor
normalized to the volume $V$ and $\phi_{i,f}({\mathbf r}_2)$ are
the relativistic wave functions of the  hydrogen atom where the
index $i$ for the initial state and the index $ f $ stands for the
final state. As we study the elastic excitation by electronic
impact, we have $ f=i=(1s_{1/2})$. The velocity of light is $
c=137.036 $ in atomic units the explicit expression of the wave
functions $\phi({\mathbf r})$ for the fundamental state (spin up)
can be found in \cite{8} and reads in atomic units as :
\begin{eqnarray}
\phi({\mathbf r})=\frac{1}{\sqrt{4\pi}}\left(\begin{array}{l}
ig(r)\\0\\
f(r)\cos(\theta)\\
f(r)\sin(\theta)e^{i\phi}
\end{array}\right)\label{2}
\end{eqnarray}
with $g(r)$  given by :
\begin{eqnarray}
g(r)=(2Z)^{\gamma+1/2}\sqrt{\frac{1+\gamma}{2\Gamma(1+2\gamma)}}e^{-Zr}r^{\gamma-1}\label{3}
\end{eqnarray}
whereas $f(r)$ is given by :
\begin{eqnarray}
f(r)=-(2Z)^{\gamma+1/2}\sqrt{\frac{1+\gamma}{2\Gamma(1+2\gamma)}}e^{-Zr}r^{\gamma-1}\left(\frac{1-\gamma}{Z\alpha}\right)\label{4}
\end{eqnarray}
To simplify the notation we shall use throughout this work the
following abbreviations :
\begin{eqnarray}
g(r)= N_ge^{-Zr}r^{\gamma -1},\quad
f(r)=-g(r)\left(\frac{1-\gamma}{Z\alpha}\right)=N_fe^{-Zr}r^{\gamma
-1}\label{5}
\end{eqnarray}
In Eq.(\ref{1}), $V_d$ is the direct interaction potential :
\begin{eqnarray}
V_d=\frac{1}{r_{12}}-\frac{Z}{r_1}\label{6}
\end{eqnarray}
where ${\mathbf r}_1$ are the electron coordinates and ${\mathbf
r}_2$ are the atomic electron coordinates and $r_{12}=|{\mathbf
r}_1-{\mathbf r}_2|$. The parameter $ \gamma $ appearing in all
these equations is :
 \begin{eqnarray}
\gamma = \sqrt{1-Z^2\alpha^2}\label{7}
\end{eqnarray}
It is straight forward to get for the transition amplitude:
 \begin{eqnarray}
S_{fi} = -i \frac{1}{\sqrt{2E_i2E_f}}\frac{1}{V}2\pi
\delta(E_f-E_i) \overline{u}(p_f,s_f)\gamma^0u(p_i,si)
H(\Delta)\label{8}
\end{eqnarray}
Where $ \gamma^0 $ is given in the standard representation of the
Dirac matrices by $ \gamma^0 = diag(1_{2},-1_2)$. The symbol $ 1_2
$ denotes the $ 2 \times 2 $ unit matrice. The argument of the
function  $ H $ is $\Delta=|{\mathbf p}_i-{\mathbf p}_f|$, the
momentum transfer. The DCS is given by:
 \begin{eqnarray}
\frac{d\sigma}{d\Omega_f}=\left.\frac{|{\mathbf p}_f|}{|{\mathbf
p}_i|}\frac{1}{(4\pi
c^2)^2}\left(\frac{1}{2}\sum_{s_i,s_f}|\ol{u}(p_f,s_f)\gamma^0u(p_i,s_i)|^2\right)|H(\Delta)|^2\right|_{E_f=E_i}\label{9}
\end{eqnarray}
In Eq.(\ref{9}), we have summed over the final polarization $s_f$
and averaged over the initial polarization $s_i$. For elastic
collisions $|{\mathbf p}_f|=|{\mathbf p}_i|=|{\mathbf p}|$ so
that $E_i=E_f=E$ and :
\begin{eqnarray}
\frac{1}{2}\sum_{s_i,s_f}|\ol{u}(p_f,s_f)\gamma^0u(p_i,s_i)|^2=4E^2\big(1-\beta^2\sin^2(\theta/2)\big)
\label{10}
\end{eqnarray}
with $\beta=|{\mathbf p}|\frac{c}{E}$. The angle $\theta$ is the
scattering angle between the vectors ${\mathbf p}_i$ and
${\mathbf p}_f$. We then have for the unpolarized DCS :
\begin{eqnarray}
\frac{d\sigma}{d\Omega_f}=\left.\frac{1}{(4\pi
c^2)^2}4E^2(1-\beta^2\sin^2(\theta/2))|H(\Delta)|^2\right|_{E_f=E_i}\label{11}
\end{eqnarray}

We now turn to the function $H(\Delta)$ of the momentum transfer
which is simply proportional to the Fourier transform of the
average (static) potential felt by the incident electron in the
field of the hydrogen atom \cite{4}. Performing the various
integrals, we get for this Fourier transform :
\begin{eqnarray}
H(\Delta)=4\pi(N_g^2+N_f^2)\Gamma(2\gamma+1)\left(\frac{1}{(2Z)^{2\gamma+1}\Delta^2}-\frac{\sin(2\gamma\phi)}{2\gamma\lambda^{2\gamma}\Delta^3}\right)\label{12}
\end{eqnarray}
where the quantities $\lambda$ and $\phi$ are :
\begin{eqnarray}
\lambda=\sqrt{(2Z)^2+\Delta^2}\quad and \quad
\phi=\arctan\left(\frac{\Delta}{2Z}\right)\label{13}
\end{eqnarray}
Even if it may not seem so, the function $H(\Delta)$ is well
behaved for the case of forward scattering $\theta=0^{\circ}$ (
recall that $\Delta=2|{\mathbf p}_i|\sin(\theta/2)$) and has the
property :
\begin{eqnarray}
\lim_{\Delta\to
0}\left(\frac{1}{(2Z)^{2\gamma+1}\Delta^2}-\frac{\sin(2\gamma\phi)}{2\gamma\lambda^{2\gamma}\Delta^3}\right)=\frac{(2\gamma+1)(2\gamma+2)}{6(2Z)^{2\gamma+3}}\label{14}
\end{eqnarray}
We must of course recover the result in the non relativistic
limit ($\beta\to 0$ and $\gamma\to 1$). In that case, the
differential cross section  is simply given by :
\begin{eqnarray}
\frac{d\sigma}{d\Omega_f}=4\frac{(\Delta^2+8)^2}{(\Delta^2+4)^4}\label{15}
\end{eqnarray}
Taking  $\beta\to 0$ and $\gamma\to 1$ (for Z=1), one easily
recovers from Eq. (\ref{11}) the above mentioned non relativistic
limit.
%**************************************************************************
\section{The differential cross section in presence of a laser field}
%***************************************************************************
We turn to the calculation of the DCS for elastic scattering
without exchange in the first Born approximation and in the
presence of a laser field. The (direct) transition amplitude in
this case is given by :
\begin{eqnarray}
S_{fi}&=&-i\int_{-\infty}^{+\infty}dt<\psi_{q_f}({\mathbf r}_1,t)\phi_f({\mathbf r}_2,t)|V_d|\psi_{q_i}({\mathbf r}_1,t)\phi_i({\mathbf r}_2,t)>\nonumber\\
&=&-\frac{i}{c}\int d^4x_1\ol{\psi}_{q_f}({\mathbf
r}_1,t)\gamma^0\psi_{q_i}({\mathbf r}_1,t)<\phi_f({\mathbf
r}_2)|V_d|\phi_i({\mathbf r}_2)>\label{16}
\end{eqnarray}
where $\phi_{i,f}({\mathbf r}_2)$ are the relativistic wave
functions of the hydrogen atom and the functions $\psi_q({\mathbf
r}_1,t)$ are the Dirac-Volkov solutions normalized to the volume
$V$ :

\begin{eqnarray}
\psi_q({\mathbf
r},t)=\frac{1}{\sqrt{2QV}}R(q)u(p,s)e^{-is(x)}\label{17}
\end{eqnarray}
with : \begin{eqnarray}
R(q)=R(p)=\left(1+\frac{1}{2(k.p)c}\ks(\as_1\cos(\varphi)+\as_2\sin(\varphi))\right)\label{18}
\end{eqnarray}
and :
\begin{eqnarray}
S(x)=q.x+\frac{(a_1.p)}{c(k.p)}\sin(\varphi)-\frac{(a_2.p)}{c(k.p)}\cos(\varphi)\label{19}
\end{eqnarray}
in the case of a circularly polarized electromagnetic potential
such that $A^{\mu}=a_1^{\mu}\cos(\varphi)+a_2^{\mu}\sin(\varphi)$
with $k_{\mu}A^{\mu}=0$ (the Lorentz condition) and
$A^2=a_1^2=a_2^2=a^2$, $a_1.a_2=0$ and $k.a_1=k.a_2=0.$ The
four-vector  $q^{\mu}=(Q/c,{\mathbf q})$ is the four-momentum of
the electron inside the laser field with wave
four-vector $k^{\mu}$.\\
We have :
\begin{eqnarray}
q^{\mu}=p^{\mu}-\frac{a^2}{2(k.p)c^2}k^{\mu}\label{20}
\end{eqnarray}
In Eq(\ref{20}) $a^2$ denotes the time-averaged square of the
four-vector potential of the laser field. The square of this
four-vector is :
\begin{eqnarray}
q_{\mu}q^{\mu}=m_*^2c^2\label{21}
\end{eqnarray}
The parameter $m_*$ plays the role of an effective mass of the
electron inside the electromagnetic field :
\begin{eqnarray}
m_*^2=1-\frac{a^2}{c^4}\label{22}
\end{eqnarray}
The factor $R(q)$ acting on the bispinor $u$ contains information
about the spin-dressing field interaction. Thus, the Dirac-Volkov
wave function represents a free-electron wave (containing a
field-dependent phase) modulated by a wave generated by the
interaction of the spin with the classical single mode field with
four-vector potential $A^{\mu}$.\\
\indent In Eqs. (\ref{18},\ref{19}),
$\varphi=k.x=k_{\mu}x^{\mu}=k_0x^0-{\mathbf k}.{\mathbf x}$ and
we use throughout this work the notations and conventions of
Bjorken and Drell \cite{8}. Proceeding along the lines of standard
calculations in QED \cite{8}, one has for the DCS :
\begin{eqnarray}
\frac{d\sigma}{d\Omega_f}=\sum_{s=-\infty}^{\infty}\frac{d\sigma^{(s)}}{d\Omega_f}\label{23}
\end{eqnarray}
where $\frac{d\sigma^{(s)}}{d\Omega_f}$ is the DCS corresponding
to the net exchange of $s$ photons and reads :
\begin{eqnarray}
\frac{d\sigma^{(s)}}{d\Omega_f}=\left.\frac{1}{(4\pi
c)^2}\frac{|{\mathbf q}_f|}{|{\mathbf
q}_i|}\left(\frac{1}{2}\sum_{s_is_f}|M_{fi}^{(s)}|^2\right)|H^{(s)}(\Delta^{(s)})|^2\right|_{Q_f=Q_i+sw}\label{24}
\end{eqnarray}
Here $\Delta_s=|{\mathbf q}_i+s{\mathbf k}-{\mathbf q}_f|$ is the
momentum transfer with the net exchange of $s$ photons. The
quantity $(\frac{1}{2}\sum_{s_is_f}|M^{(s)}_{fi}|^2)$ is the
electronic contribution to the differential cross section
$\frac{d\sigma^{(s)}}{d\Omega}$ and has already been determined
in a previous work \cite{9}. We simply quote the final result:
\begin{eqnarray}
&&\frac{1}{2}\sum_{s_is_f}|M^{(s)}_{fi}|^2\nonumber\\&=&\frac{2}{c^2}\{J_s^2(z)A+
\big(J^2_{s+1}(z)+J^2_{s-1}(z)\big)B\nonumber\\
&+&\big(J_{s+1}(z)J_{s-1}(z)\big)C+J_s(z)\big(J_{s-1}(z)+J_{s+1}(z)\big)D\}\label{25}
\end{eqnarray}
where the argument $z$ of the Bessel functions is given by
$z=\sqrt{\alpha_1^2+\alpha_2^2}$ and $\alpha_1$ and $\alpha_2$
are such that :
\begin{equation}
\alpha_1=\frac{1}{c}\left\{\frac{a_1.p_i}{k.p_i}-\frac{a_1.p_f}{k.p_f}\right\},\quad
\alpha_2=\frac{1}{c}\left\{\frac{a_2.p_i}{k.p_i}-\frac{a_2.p_f}{k.p_f}\right\}\label{26}
\end{equation}
The coefficients $A$, $B$, $C$ and $D$ are respectively given by :
\begin{eqnarray}
A&=&c^4-(q_f.q_i)c^2+2Q_fQ_i-\frac{a^2}{2}\left(\frac{(k.q_f)}{(k.q_i)}+\frac{(k.q_i)}{(k.q_f)}\right)+\frac{a^2\omega^2}{c^2(k.q_f)(k.q_i)}((q_f.q_i)-c^2)+\nonumber\\
&&\frac{(a^2)^2\omega^2}{c^4(k.q_f)(k.q_i)}+\frac{a^2\omega}{c^2}(Q_f-Q_i)\left(\frac{1}{(k.q_i)}-\frac{1}{(k.q_f)}\right)\label{27}
\end{eqnarray}
\begin{eqnarray}
B&=&-\frac{(a^2)^2\omega^2}{2c^4(k.q_f)(k.q_i)}+\frac{\omega^2}{2c^2}\left(\frac{(a_1.q_f)}{(k.q_f)}\frac{(a_1.q_i)}{(k.q_i)}+\frac{(a_2.q_f)}{(k.q_f)}\frac{(a_2.q_i)}{(k.q_i)}\right)-\frac{a^2}{2}+\nonumber\\
&&\frac{a^2}{4}(\frac{(k.q_f)}{(k.q_i)}+\frac{(k.q_i)}{(k.q_f)})-\frac{a^2\omega^2}{2c^2(k.q_f)(k.q_i)}\big((q_f.q_i)-c^2\big)+\nonumber\\
&&\frac{a^2\omega}{2c^2}(Q_f-Q_i)\left(\frac{1}{(k.q_f)}-\frac{1}{(k.q_i)}\right)\label{28}
\end{eqnarray}
\begin{eqnarray}
C&=&\frac{\omega^2}{c^2(k.q_f)(k.q_i)}\big(\cos(2\phi_0)\{(a_1.q_f)(a_1.q_i)-(a_2.q_f)(a_2.q_i)\}+\nonumber\\
&&\sin(2\phi_0)\{(a_1.q_f)(a_2.q_i)+(a_1.q_i)(a_2.q_f)\}\big)\label{29}
\end{eqnarray}
\begin{eqnarray}
D&=&\frac{c}{2}\left((\AA.q_i)+(\AA.q_f)\right)-\frac{c}{2}\left(\frac{(k.q_f)}{(k.q_i)}(\AA.q_i)+\frac{(k.q_i)}{(k.q_f)}(\AA.q_f)\right)+\nonumber\\
&&\frac{\omega}{c}\left(\frac{Q_i(\AA.q_f)}{(k.q_f)}+\frac{Q_f(\AA.q_i)}{(k.q_i)}\right)\label{30}
\end{eqnarray}
The function $H^{(s)}(\Delta_s)$ is now given by :
\begin{eqnarray}
H^{(s)}(\Delta_s)=4\pi(N_g^2+N_f^2)\Gamma(2\gamma+1)\left(\frac{1}{(2Z)^{2\gamma+1}(\Delta_s)^2}-\frac{\sin(2\gamma\phi_s)}{2\gamma\lambda_s^{(2\gamma)}(\Delta_s)^3}\right)\label{31}
\end{eqnarray}
with $\lambda_s=\sqrt{(2Z)^2+(\Delta_s)^2}$ and
$\phi_s=\arctan(\Delta_s/2Z)$. Once again, the function
$H^{(s)}(\Delta_s)$ is well behaved for the forward scattering
witch corresponds to $s=0$. When no radiation field is present,
all Bessel functions vanish except for $s=0$: We have
$J_s(z=0)=\delta_{s0}$ and in this case, the result reduces to
the unpolarized DCS given in Eq. (11)
%***********************************************************************************
\section{Results and discussions}
%*********************************************************************************
The kinematics of the process is that given in \cite{9} and we
maintain the same choice for the laser angular frequency, that is
$w=0.043$ which corresponds to a near-infrared neodymium laser.

%***********************************************************************************
\subsection{The Nonrelativistic regime}
%***********************************************************************************
In the limit of low electron kinetic energy and moderate field
strength, typically an electron kinetic energy $W=100\;a.u$ and a
field strength $\varepsilon =0.05\; u.a$, it is easy to see that
the effects of the additional spin terms and the dependence of
$q^{\mu}$ on the spatial orientation of the electron momentum due
to $(k.p)$ are expected to be small. For $W=100\;a.u$ and
$\varepsilon =0.05\; u.a$, some angles $\theta({\mathbf
q}_i,{\mathbf q}_f)$ and $\theta({\mathbf p}_i,{\mathbf p}_f)$
are given in Table.1 and this shows indeed that the dependence of
$q^{\mu}$ (and hence of ${\mathbf q}$) on the spatial orientation
of the electron momentum is indeed small in the nonrelativistic
regime.\\
\indent In Fig. (1), we compare the non relativistic DCS given by
Eq. (\ref{15}) and the relativistic DCS given by Eq. (\ref{11}) as
functions of the scattering angle $\theta({\mathbf p}_i,{\mathbf
p}_f)$ in absence of the laser field. As expected, in the
nonrelativistic limit, there is only a small difference between
these two cross sections and we see that this difference becomes
more pronounced for the case of forward scattering. For large
angle scattering, there is almost no difference between the
nonrelativistic calculations and the relativistic calculations. In
Fig. (2), instead of plotting
$\left(\frac{d\sigma}{d\Omega_f}\right)_{NR}$ and
$\left(\frac{d\sigma}{d\Omega_f}\right)_{REL}$ as functions of
the scattering angle, we use the angular coordinates
$(\theta_i,\phi_i)$ of ${\mathbf p}_i$ and $(\theta_f,\phi_f)$ of
${\mathbf p}_f$ to plot the angular dependence of the two DCSs as
functions of $\theta_f$, the angle between ${\mathbf p}_f$ and
the $Oz$ axis. This will serve as a consistency check of our next
calculations in presence of an electromagnetic potential
circularly polarized  and whose wave-vector points in the $Oz$
direction. We have chosen a geometry where
$\theta_i=\phi_i=45^{\circ}$ and the angle $\theta_f$ varies from
$0^{\circ}$ to $180^{\circ}$ with $\phi_f=90^{\circ}$. The
relativistic parameter $\gamma_{rel}=1/\sqrt{1-\beta^2}$ is equal
to $1.0053$ which corresponds to an electron kinetic energy equal
to $100\;u.a\simeq 2.721\;kev$. The first observation to be made
is that in the non relativistic regime, the non relativistic DCS
is very close to the relativistic one, which was to be expected.
Also, there a peak in the vicinity of $\theta_f=35^{\circ}$.\\
\indent In Fig. (3), we compare the relativistic summed DCS with
and without laser field for the net exchange of $\pm 100$
photons. As one can see, the laser field gives rise to important
modifications of the DCS. for this collision geometry, several
hundred photons can be exchanged even in the case of a moderate
laser intensity of $8.75\times 10^{13}\,W/cm^2$. In Fig. (4), the
net exchange of $\pm 200$ photons shows that the DCS with laser
field approaches the DCS without laser field and we have almost
two indistinguishable curves in the case of the net exchange $\pm
300$ photons as one can see in Fig. (5).\\
\indent It is also important to compare the relativistic DCS
$\left(\frac{d\sigma^{(s)}}{d\Omega}\right)$ corresponding to the
net exchange of $s$ photons where only the electronic dressing
term is taken into account, with the corresponding non
relativistic DCS. Working with the nonrelativistic Volkov states,
one easily gets for the non relativistic case :
\begin{equation}
\frac{d\sigma^{B_1,s}}{d\Omega_f}=\frac{|{\mathbf
p}_f(s)|}{|{\mathbf p}_i|}J_s^2\left(\frac{|A|}{cw}|{\mathbf
p}_i-{\mathbf p}_f(s)|\right)\times
\frac{d\sigma^{B_1,F,F}}{d\Omega_f}\label{32}
\end{equation}
where the first-Born DCS is given by Eq. (\ref{15}) and
corresponds to the field free case. In Fig. (6), we compare the
non relativistic and relativistic DCS for an exchange of $\pm
100)$ photons and for a geometry where
$\theta_i=\phi_i=45^{\circ}$ and $\phi_f=90^{\circ}$. In the non
relativistic regime, both approaches give similar results
particularly for large angle scattering namely $\theta_f\geq
45^{\circ}$. In Fig. (7), and for the same values of the angular
parameters, one cannot distinguish the two curves in the case of
an exchange of $\pm 300$ photons above $70^{\circ}$  .Beyond this
value, the non relativistic DCS is lower than the relativistic
one. In Fig. (8), the DCSs for the absorption of one photon are
shown and there is a qualitative difference between the two
approaches. Indeed, we have already mentioned that there is a
peak in the vicinity of $\theta_f= 35^{\circ}$ when we have
plotted the two DCS as functions of $\theta_f$. This peak is
peculiar to the geometry chosen and the relativistic DCS
corresponding to the absorption of one photon is peaked for
$\theta_f\simeq 35^{\circ}$ whereas the non relativistic one is
peaked at about $\theta_f = 30^{\circ}$. One explanation that can
be given to this difference is that spin-effects as well as
relativistic electronic dressing that are fully taken into
account in the relativistic treatment influence the location of
such peaks. This is another reason that gives a strong motivation
to study relativistic effects in laser-assisted electron-atom
collisions. Experimentalists need to have theoretical predictions
that will enable them to focus on a particular direction. Of
course, this is only a modification. due to the relativistic
electronic dressing. To be complete, we shall present very soon,
a treatment that takes into account the relativistic dressing of
the atomic system by the strong radiation field. In Fig. (9), and
for the same values of angular parameters, the DCSs for the
emission of one photon are given and as expected there is a
complete analogy between the process of absorption and emission.
In fact, these two processes give the same values for the DCSs.

%***********************************************************************************************
\subsection{The relativistic regime}
%**************************************************************************************************
In the limit of high electron kinetic energy and strong field
strength, typically an electron kinetic energy $W=c^2\; a.u$ and a
field strength $\varepsilon=1.00\;a.u$, the effects of the
additional spin terms and the dependence of $q^{\mu}$ on the
spatial orientation of the electron momentum due to $(k.p)$ begin
top be noticeable. For these values of $W$ and $\varepsilon$, some
$\cos$ of angle $\theta({\mathbf q}_i,{\mathbf q}_f)$ and
$\theta({\mathbf p}_i,{\mathbf p}_f)$ are given in Table.2 and
this shows  indeed that the dependence of $q^{\mu}$ (and hence of
${\mathbf q}$) on the spatial orientation of the electron
momentum is not negliegeable in the relativistic regime.

%******************************************************************************************************
In Fig. (10), the non relativistic DCS is compared to the
relativistic DCS as functions of the scattering angle $\theta$
(not to be confused with the angle $\theta_f$). The
nonrelativistic formalism is no longer applicable since there is
now a net difference between the DCS given by Eq. (\ref{11}) and
the DCS given by Eq. (\ref{15}) particularly for small angles.
Indeed, for the case of forward scattering,
$\left(\frac{d\sigma}{d\Omega}\right)_{NR}=1$ while
$\left(\frac{d\sigma}{d\Omega}\right)_{R}\simeq 4$ and the
difference between the DCS remains noticeable up to
$\theta=1.5^{\circ}$. For large angles, both approaches give
nearly the same results. In Fig. (11), we compare the non
relativistic DCS and the relativistic DCS without laser field as
functions of the angle $\theta_f$. Again, these two DCS present
both a peak in the vicinity of $\theta_f\simeq 35^{\circ}$. It is
clear that the non relativistic formalism is no longer applicable
as we have considered now a relativistic parameter
$\gamma_{rel}=2$, with corresponds to an electron kinetic energy
of about 0.511 $Mev$. It is a fully relativistic regime  and we
show in Fig. (12) the DCS with (dashed line ) and without (dotted
line ) laser field for the net exchange of $\pm 3500$ photons.
Even in this case, the DCS with laser field is almost smeared out
and one has to go up to a very large number of photon exchange to
recover the relativistic DCS without laser field. Unfortunately,
and due to a lack of high speed computing facilities, we cannot
present for the time being a complete analysis of the
relativistic and the ultrarelativistic regime. However this will
be done in forthcoming paper. Now, we compare the DCSs in the
relativistic regime, for the relativistic parameter $\gamma=2$.
In Fig. (13), we compare the summed DCS relativistic and non
relativistic where there is an exchange of $\pm 500$ photons. The
values of the nonrelativistic DCS is almost halved with regard to
the relativistic DCS. In Fig (14), we show the angular
distribution of the two DCSs for the process of the absorption of
one photon. For the relativistic DCS, there is an appearance of
three peaks one located in the vicinity of
$\theta_f=22.5^{\circ}$ and double peak in the interval
$(40^{\circ},45^{\circ})$ while for the non relativistic DCS,
there is a double peak in the vicinity of $33^{\circ}$. We also
have the same qualitative result that the non relativistic DCS is
almost halved with regard to the relativistic DCS for the process
of absorption of one photon. In Fig. (15), we present the same
result but for the process of emission of one photon. The values
of the DCS are exactly the same as that of the absorption
process.\\
 Finally, in Fig. (16), we give the envelope of the
differential cross section as a function of the energy transfer
$Q_f-Q_i$ for the non relativistic regime. The final
energy-spectrum is not symmetric around  the elastic peak.
%****************************************************************************
 \section{Conclusion}
%*******************************************************************************
In this work, we have studied the effect of the relativistic
electronic dressing in laser-assisted electron-hydrogen elastic
collisions. To our knowledge, this is the first time that such a
calculation has been carried out at this level. Even if the
formalism may seem heavy and complicated, we have (using the case
with no laser field as a guide ) checked every step of our
calculations. What emerges is that the relativistic electronic
dressing reduces considerably the magnitude of the DCS. We must
sum the DCS given by Eq. (\ref{23})over a very large number of
photon in order to get the same order of magnitude as that of the
DCS given by Eq. (\ref{11}). Of course, a more sophisticated
approach is needed in order to have a complete treatment of this
relativistic process. The relativistic generalization of the
method due to F.W. Byron and C.J. Joachain \cite{10} which takes
into account the atomic dressing will be presented in a separated
paper.
 These authors and others
\cite{11} have show that in the non relativistic regime, the
dressing of atomic states can give rise to very important
modifications of the DCS. The case of inelastic collisions will
be published very soon. The non relativistic treatment of
laser-assisted electron atom collisions taking into account both
the electronic dressing and the atomic dressing has been studied
by many authors \cite{12}. All agree that at least in the
nonrelativistic regime, the effects of atomic dressing can modify
the behaviour of the DCS. The study of the relativistic
electronic dressing in laser-assisted inelastic collisions will be
sent very soon to the Los Alamos web site xxx.lanl.org.
 Recently, some authors (\cite{13},\cite{14}) have used the
strong field approximation (SFA)to study the photoionization of
hydrogen by electron impact. Work is in progress to present a
more rigorous
 approach.
 \section*{Acknowledgments}
One of us (Y. Attaourti) dedicates the present work to his
Teacher and great Master, Prof C.J. Joachain (Universit\'e Libre
de Bruxelles, Belgium),
 one of the most outstanding physicist in the field of atomic collision theory who has recently
retired and to Prof F. Brouillard (Universit\'e Catholique de
Louvain, Belgium) , a down-to-earth physicist and a very kind
human being.

\newpage

\begin{figure}
\begin{center}
\includegraphics[angle=0,width=15cm]{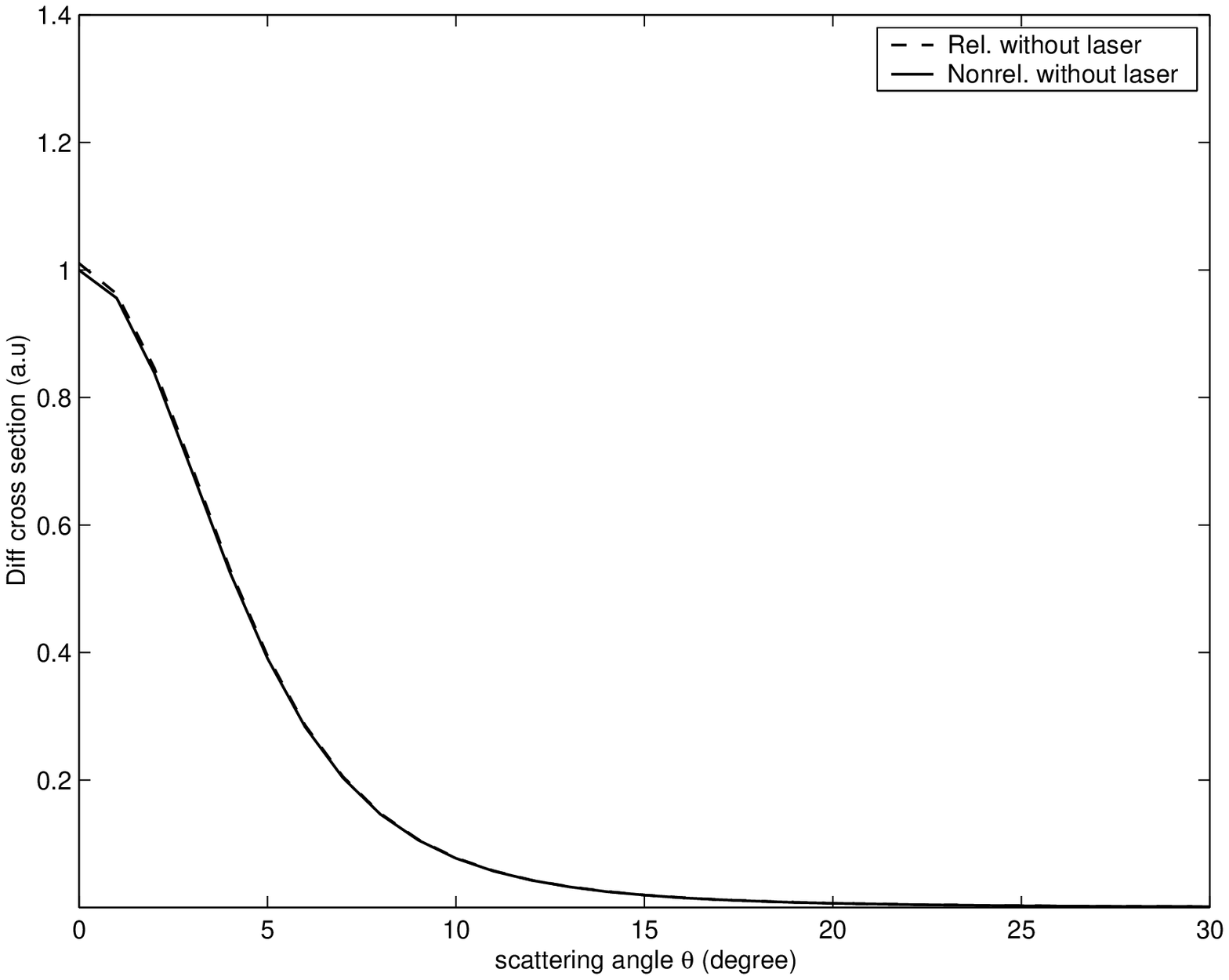}
\caption{Comparison between the nonrelativistic DCS and the
relativistic DCS as functions of the scattering angle $(\theta)$
varying from $0^{\circ}$ to $30^{\circ}$ , without taking into
account the spherical coordinates  of $(p_i,p_f)$; the curves are
perfectly confounded }
\end{center}
\end{figure}
\newpage
\begin{figure}
\begin{center}
\includegraphics[angle=0,width=15cm]{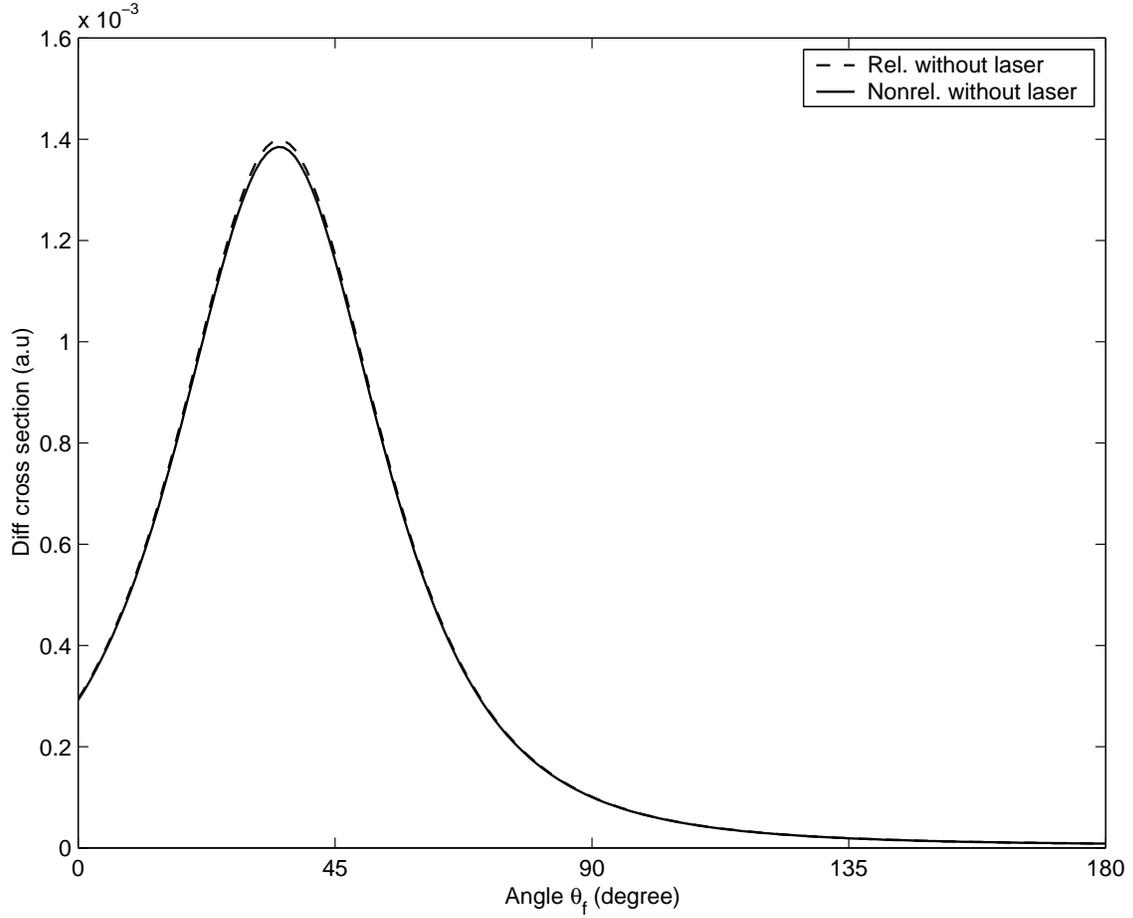}
\caption{Comparison in the nonrelativistic regime, between the
nonrelativistic DCS and the relativistic DCS as functions of
$(\theta_f)$ the angle between ${\mathbf p}_f$ and the $Oz$ axis.
The parameters are $\gamma=1.0053\,a.u$, $\varepsilon=0.05\,a.u$
and $w=0.043\,a.u$. The geometry chosen is
$\theta_i=\phi_i=45^{\circ}$ where $\theta_f$ varying from
$0^{\circ}$ to $180^{\circ}$ with $\phi_f=90^{\circ}$ }
\end{center}
\end{figure}
\newpage

\begin{figure}
\begin{center}
\includegraphics[angle=0,width=15cm]{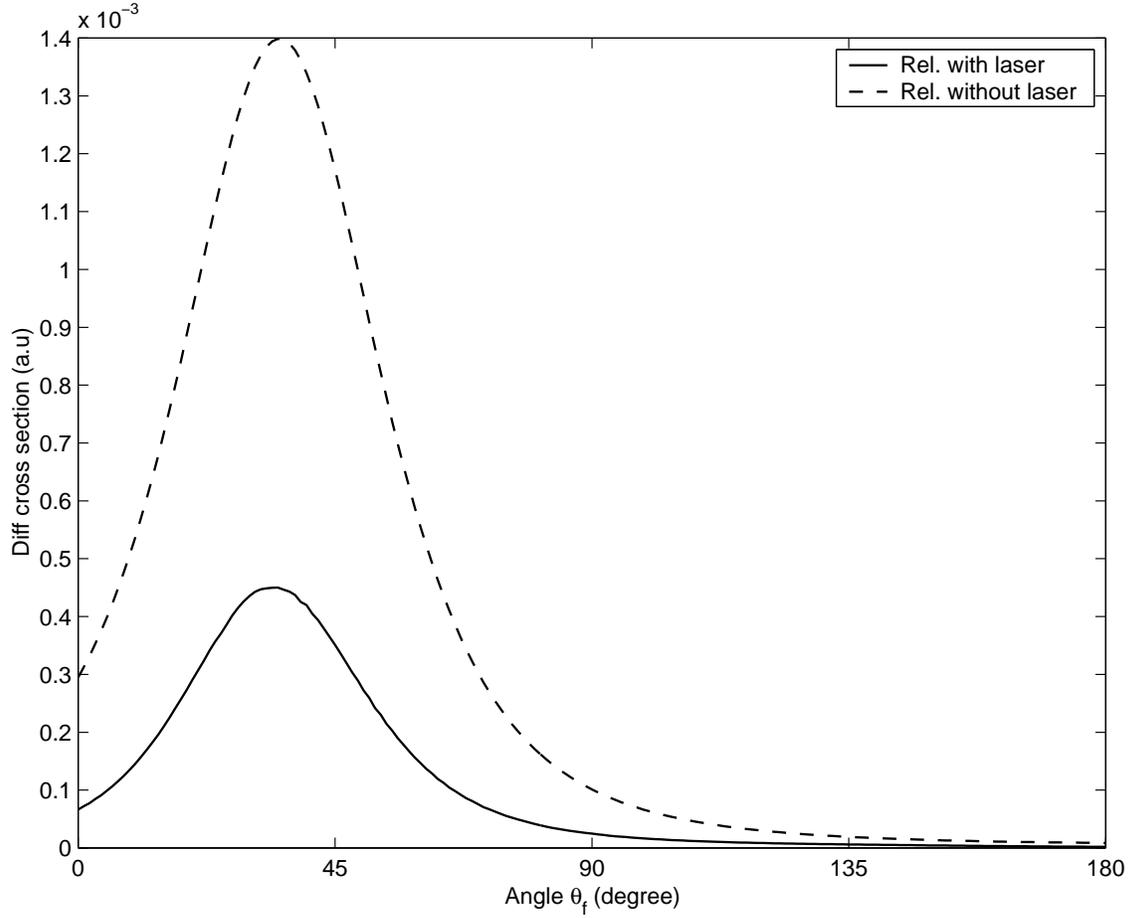}
\caption{Comparison in the nonrelativistic regime, between the
relativistic DCS with laser and the relativistic DCS without
laser for the exchange of $\pm 100$ photons. The parameters are
$\gamma=1.0053\,a.u$, $\varepsilon=0.05\,a.u$ and $w=0.043\,a.u$.
The geometry chosen is $\theta_i=\phi_i=45^{\circ}$ where
$\theta_f$ varying from $0^{\circ}$ to $180^{\circ}$ with
$\phi_f=90^{\circ}$ }
\end{center}
\end{figure}
\newpage
\begin{figure}
\begin{center}
\includegraphics[angle=0,width=15cm]{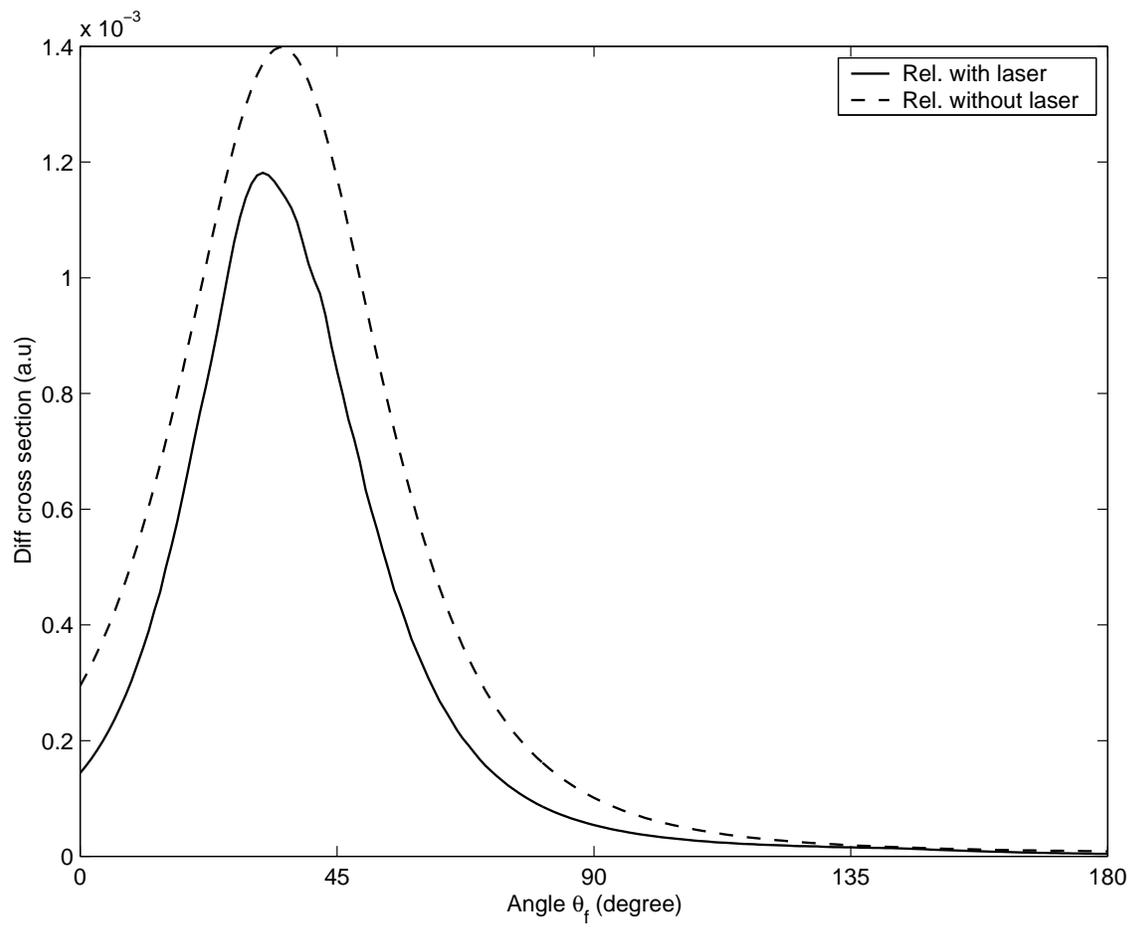}
\caption{Same as Fig. 3 but for the exchange of $\pm 200$
photons.}
\end{center}
\end{figure}

\newpage

\begin{figure}
\begin{center}
\includegraphics[angle=0,width=15cm]{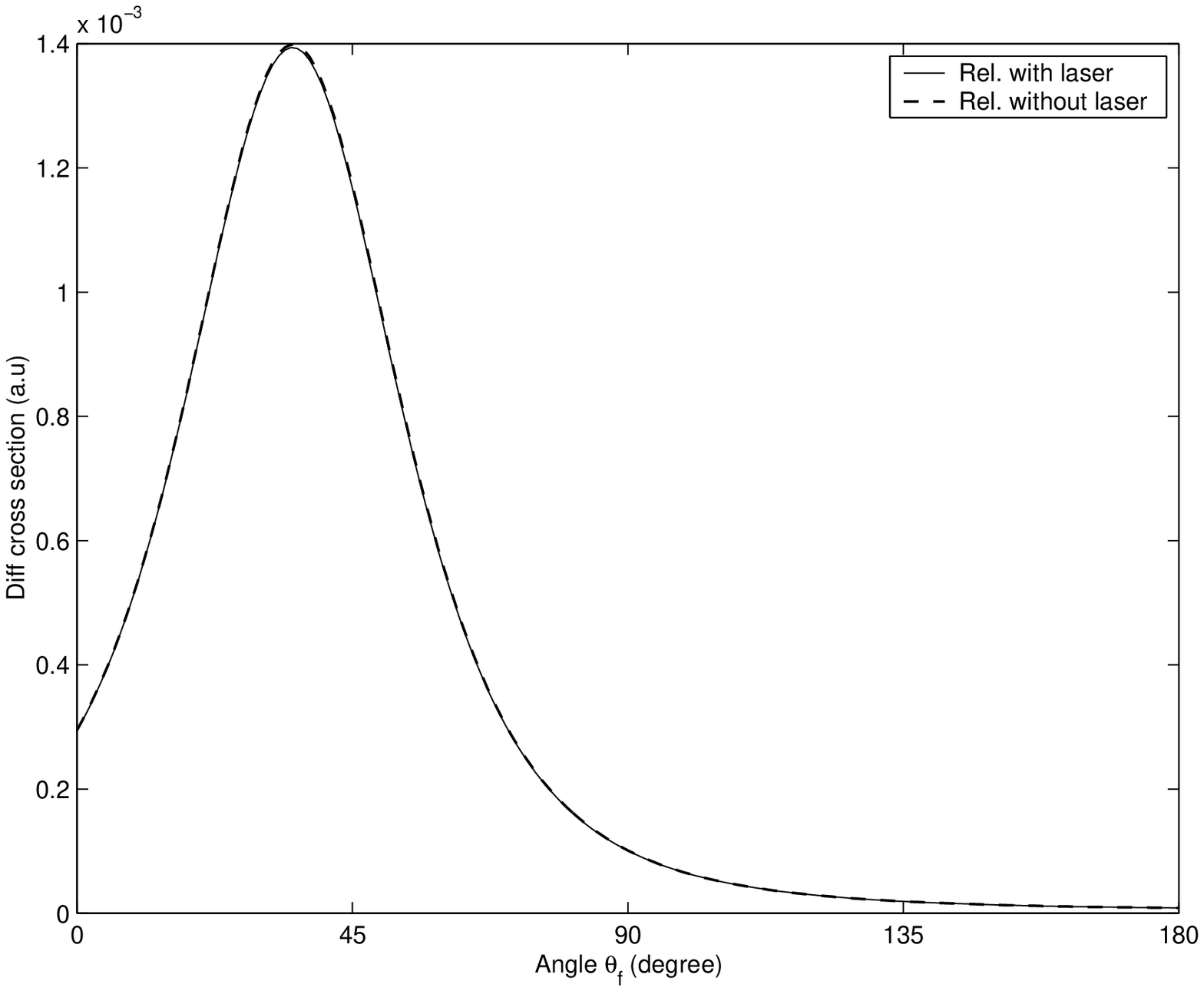}
\caption{ Same as the last figure but for an exchange of $\pm
300$ photons. The curves are indistinguishable.}
\end{center}
\end{figure}

\newpage

\begin{figure}
\begin{center}
\includegraphics[angle=0,width=15cm]{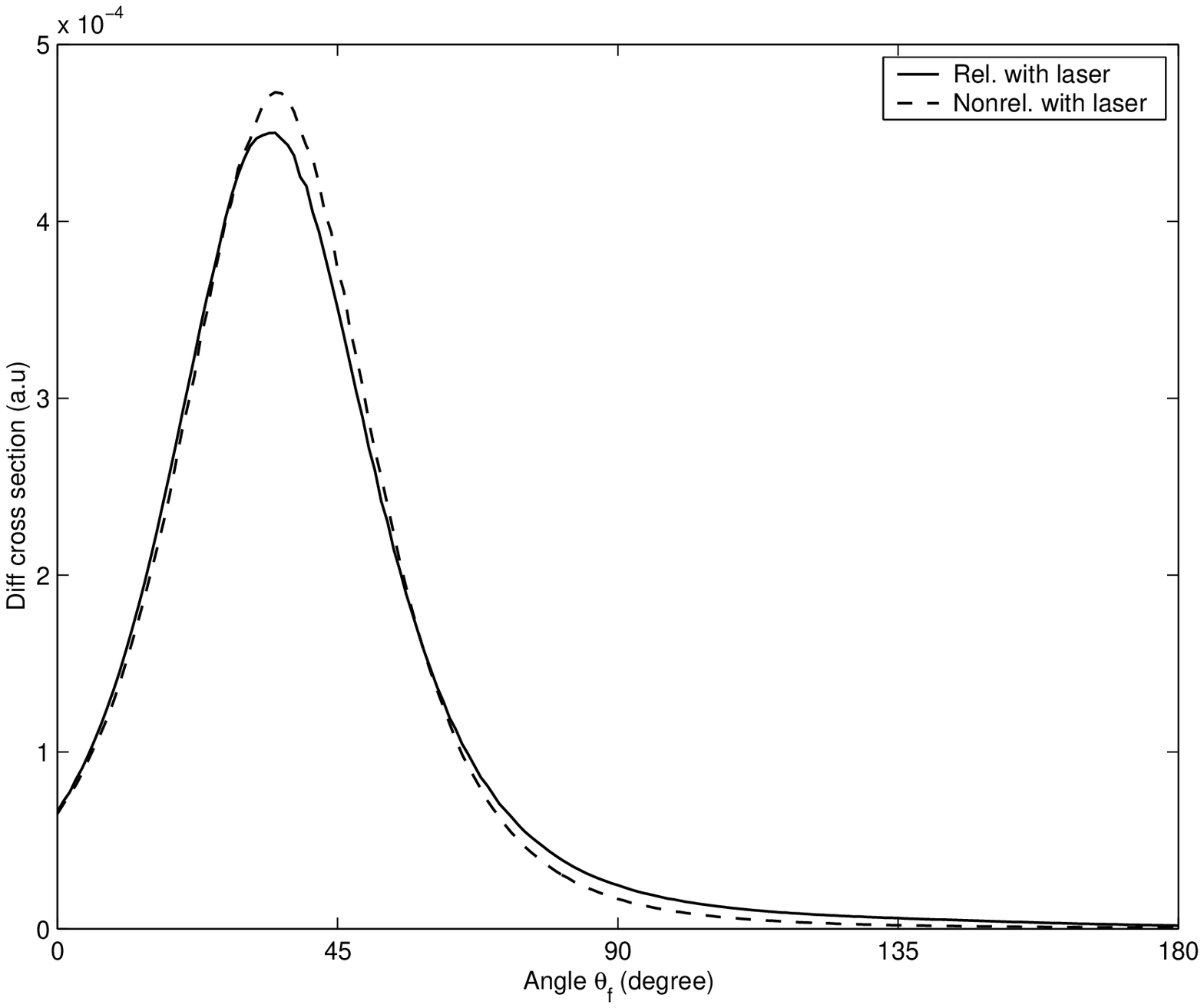}
\caption{Comparison between the relativistic DCS with laser and
the nonrelativistic DCS with laser for an exchange of $\pm100$
photons. The parameters are $\gamma=1.0053\,a.u$,
$\varepsilon=0.05\,a.u$, $w=0.043\,a.u$. The geometry chosen is
$\theta_i=\phi_i=45^{\circ}$ where $\theta_f$ varying from
$0^{\circ}$ to $180^{\circ}$ with $\phi_f=90^{\circ}$ }
\end{center}
\end{figure}

\newpage

\begin{figure}
\begin{center}
\includegraphics[angle=0,width=15cm]{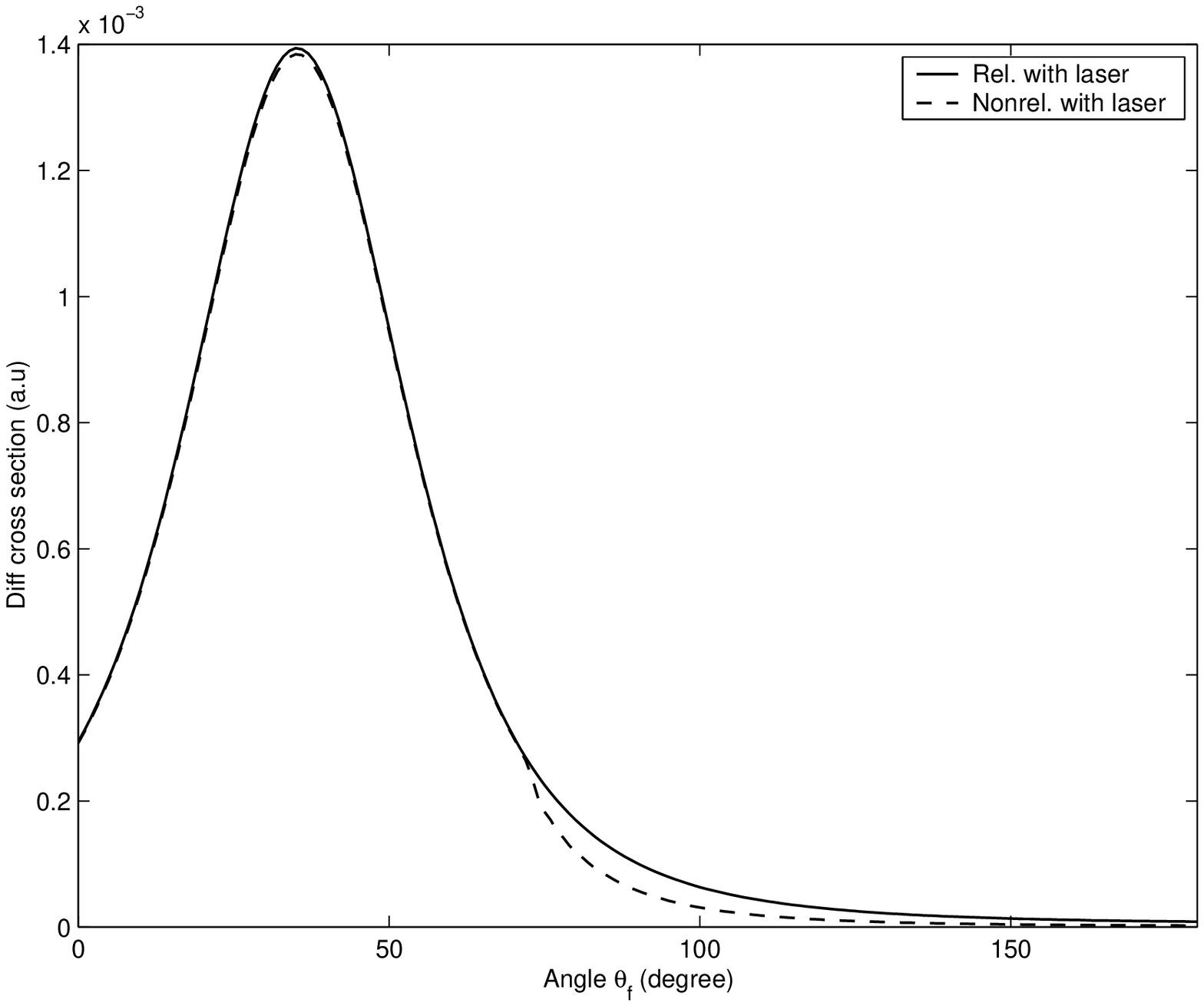}
\caption{ Same as the last Figure but for an exchange of $\pm 300$
photons. The curves are nearly indistinguishable }
\end{center}
\end{figure}
\newpage
\begin{figure}
\begin{center}
\includegraphics[angle=0,width=16cm]{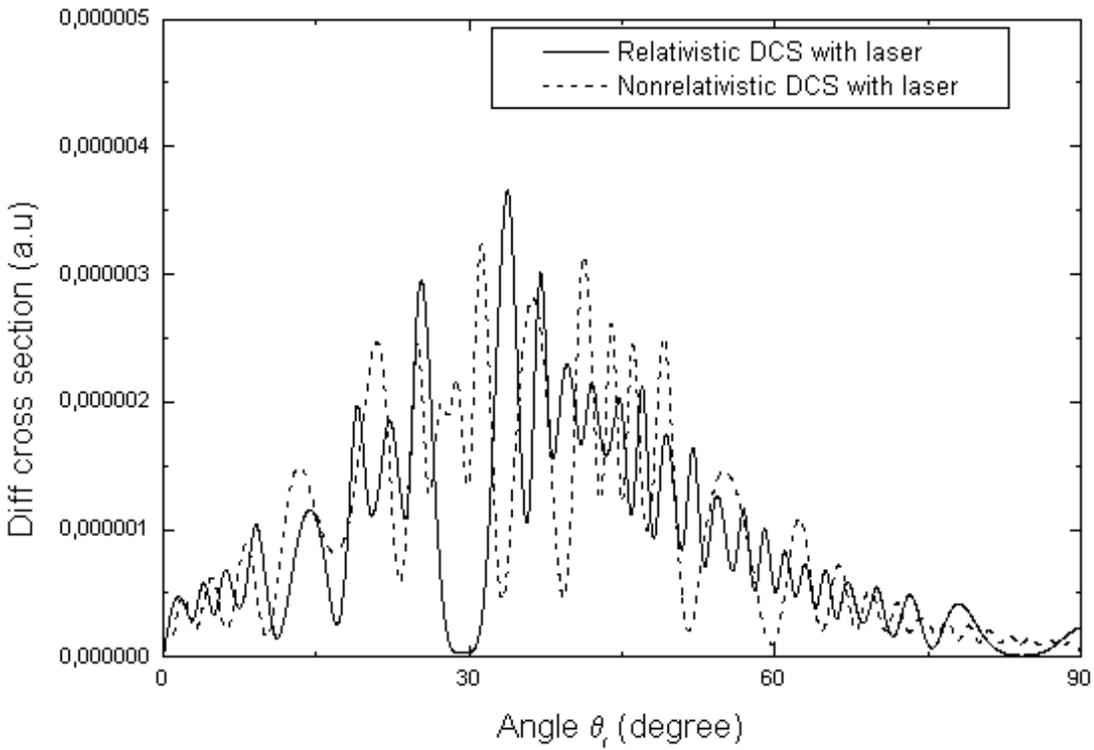}
\caption{Comparison in the nonrelativistic regime between the
relativistic DCS with laser and the nonrelativistic DCS with
laser for absorption of one photon. The geometry chosen is
$\theta_i=\phi_i=45^{\circ}$ where $\theta_f$ varying from
$0^{\circ}$ to $90^{\circ}$ with $\phi_f=90^{\circ}$. The
parameters are ($\gamma=1.0053\,a.u$, $\varepsilon=0.05\,a.u$ and
$w=0.043\,a.u$)}
\end{center}
\end{figure}
\newpage
\begin{figure}
\begin{center}
\includegraphics[angle=0,width=16cm]{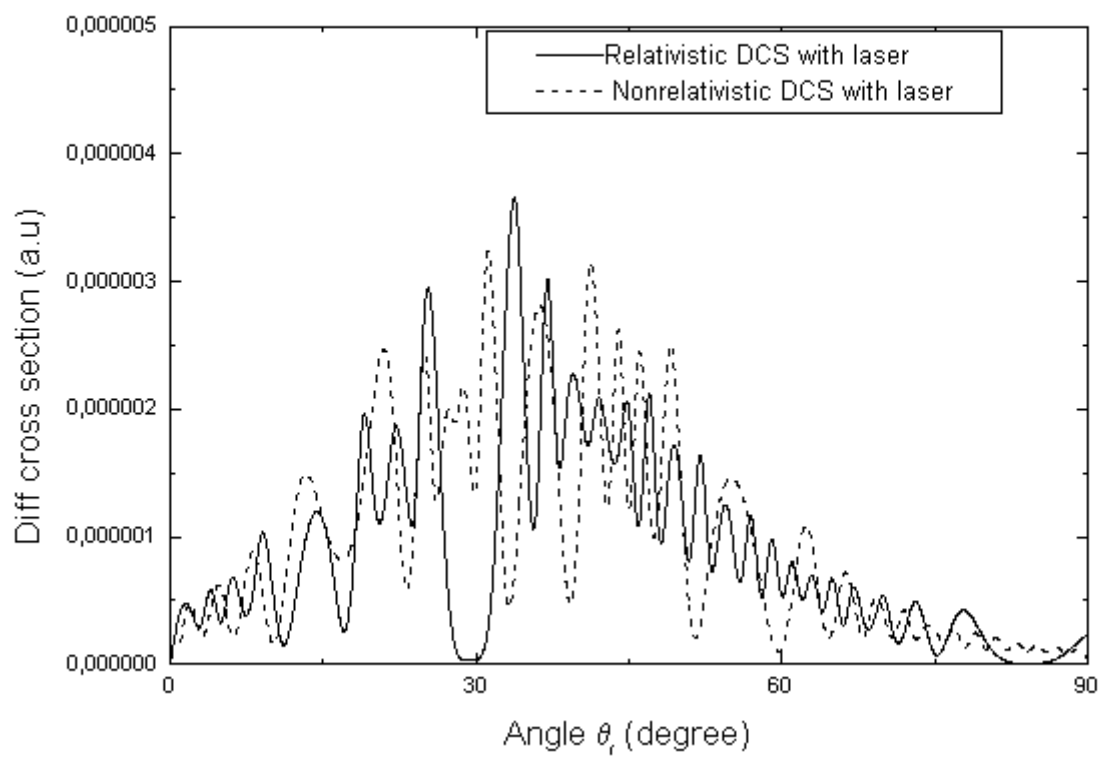}
\caption{Comparison between the relativistic DCS with laser and
the nonrelativistic DCS with laser for the emission of one photon,
The parameters are $\gamma=1.0053\,a.u$, $\varepsilon=0.05\,a.u$,
$w=0.043\,a.u$. The geometry chosen is
$\theta_i=\phi_i=45^{\circ}$ where $\theta_f$ varying from
$0^{\circ}$ to $90^{\circ}$ with $\phi_f=90^{\circ}$}
\end{center}
\end{figure}
\newpage
\begin{figure}
\begin{center}
\includegraphics[angle=0,width=15cm]{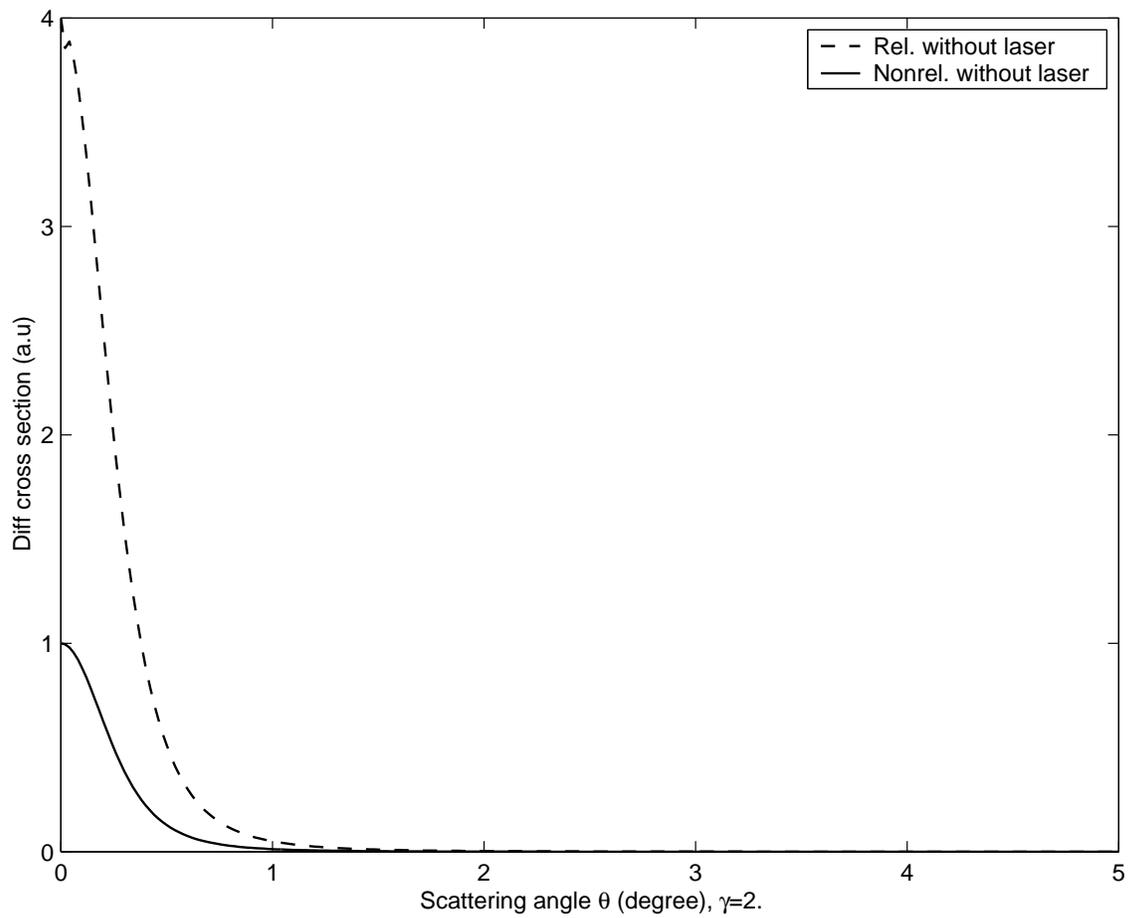}
\caption{Comparison between the nonrelativistic DCS and the
relativistic DCS as functions of the scattering angle $(\theta)$
varying from $0^{\circ}$ to $5^{\circ}$}
\end{center}
\end{figure}
\newpage
\begin{figure}
\begin{center}
\includegraphics[angle=0,width=15cm]{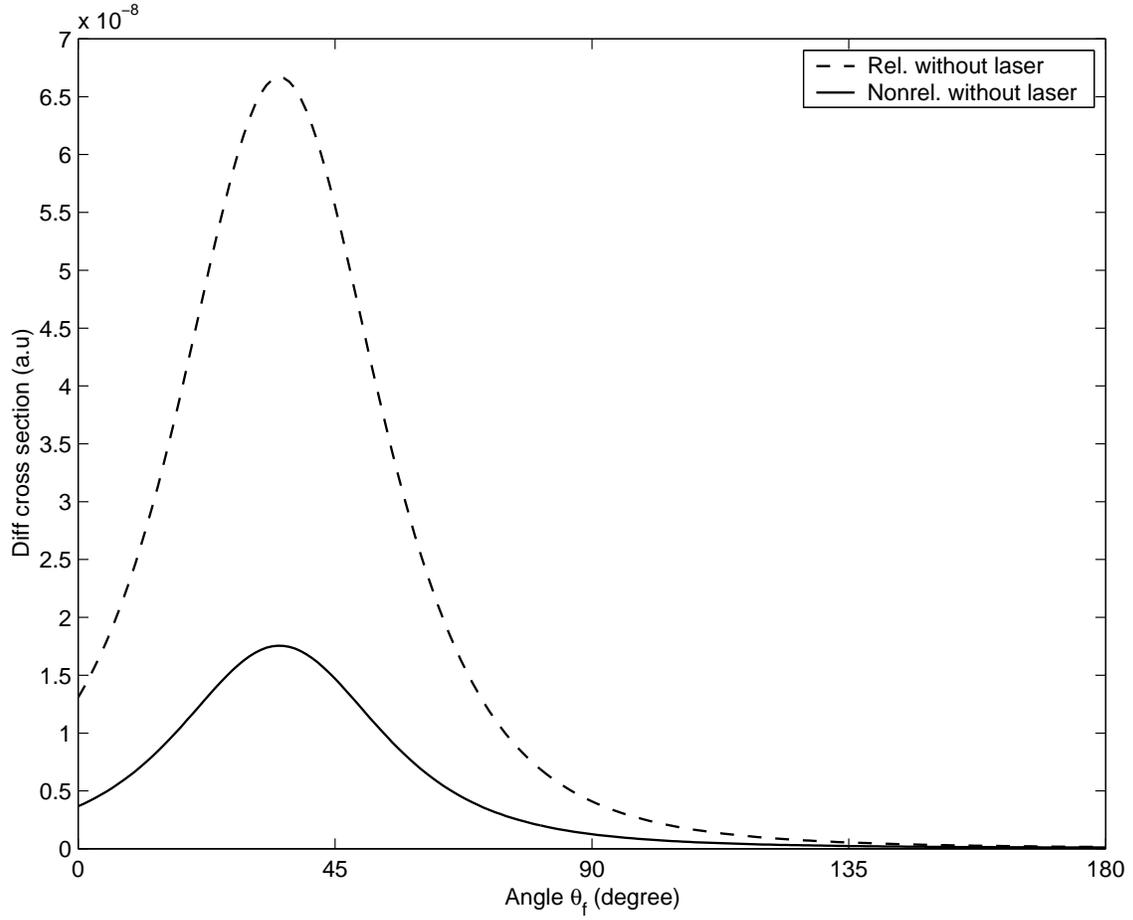}
\caption{Comparison in the relativistic regime ($\gamma=2\,a.u$,
$\varepsilon=1\,a.u$ and $w=0.043\,a.u$)  between the
nonrelativistic DCS and the relativistic DCS as functions of
$(\theta_f)$ the angle between ${\mathbf p}_f$ and the $Oz$ axis.
The geometry chosen is $\theta_i=\phi_i=45^{\circ}$ where
$\theta_f$ varies from $0^{\circ}$ to $180^{\circ}$ with
$\phi_f=90^{\circ}$ }
\end{center}
\end{figure}
\newpage
\begin{figure}
\begin{center}
\includegraphics[angle=0,width=15cm]{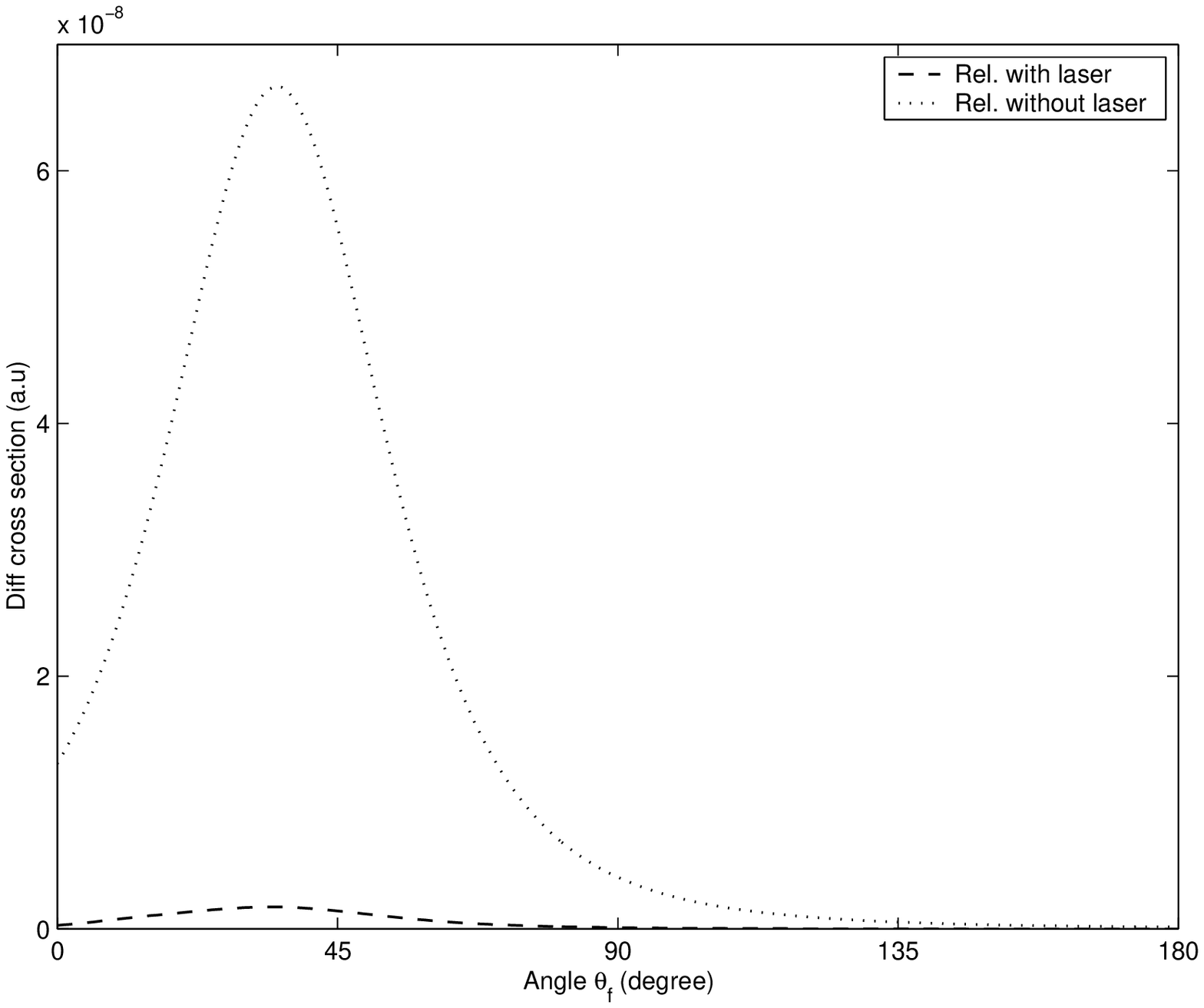}
\caption{Comparison in the relativistic regime ($\gamma=2\,a.u$,
$\varepsilon=1\,a.u$ and $w=0.043\,a.u$) between the relativistic
DCS with laser and the relativistic DCS without laser for the
exchange of $\pm 3500$ photons. The geometry chosen is
$\theta_i=\phi_i=45^{\circ}$ where $\theta_f$ varies from
$0^{\circ}$ to $180^{\circ}$ with $\phi_f=90^{\circ}$}
\end{center}
\end{figure}
\newpage
\begin{figure}
\begin{center}
\includegraphics[angle=0,width=15cm]{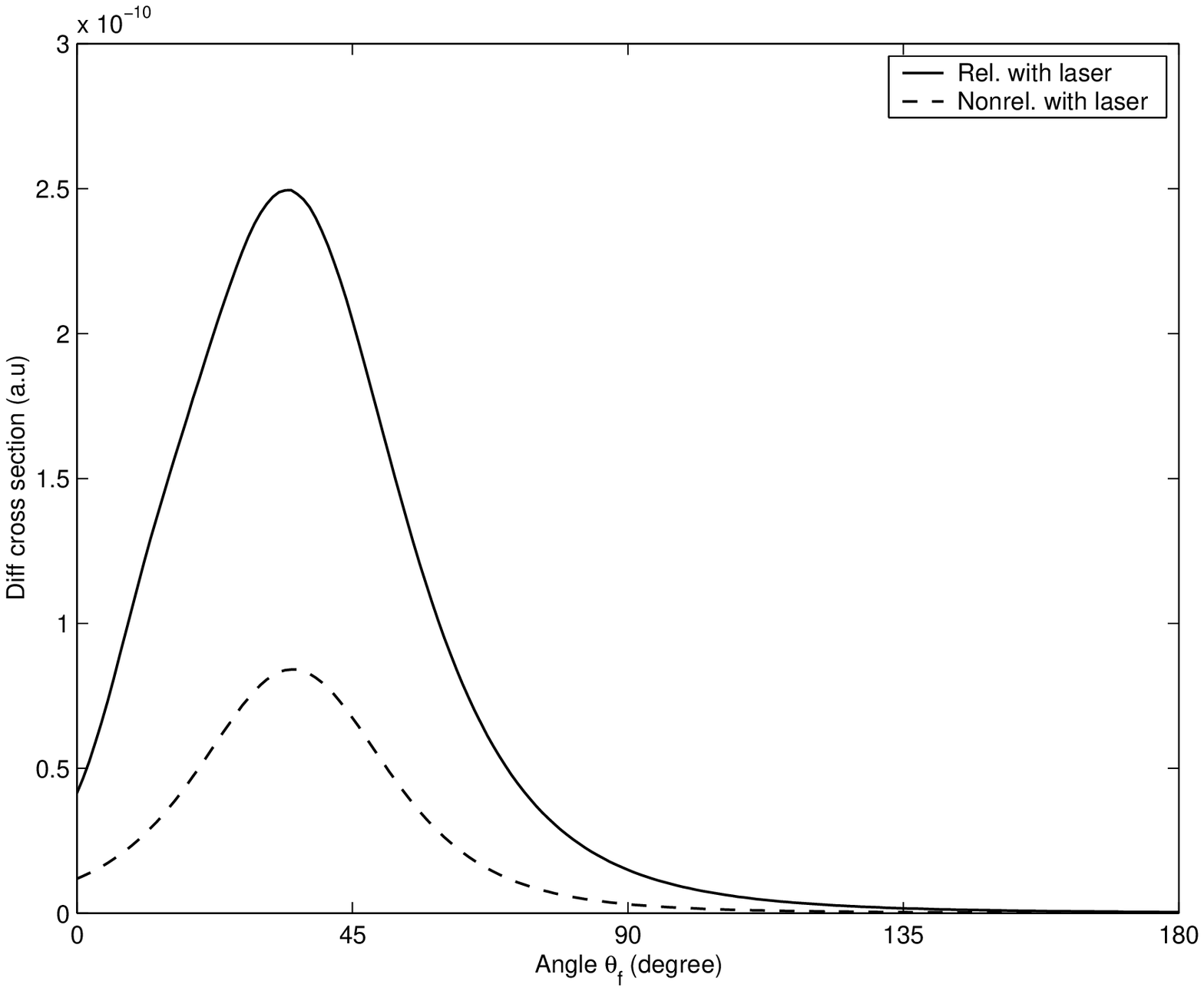}
\caption{Comparison in the relativistic regime ($\gamma=2\,a.u$,
$\varepsilon=1\,a.u$, $w=0.043\,a.u$) between the relativistic
DCS with laser and the nonrelativistic DCS with laser for an
exchange of $\pm 500$ photons. The geometry chosen is
$\theta_i=\phi_i=45^{\circ}$ where $\theta_f$ varies from
$0^{\circ}$ to $180^{\circ}$ with $\phi_f=90^{\circ}$}
\end{center}
\end{figure}
\newpage
\begin{figure}
\begin{center}
\includegraphics[angle=0,width=16cm]{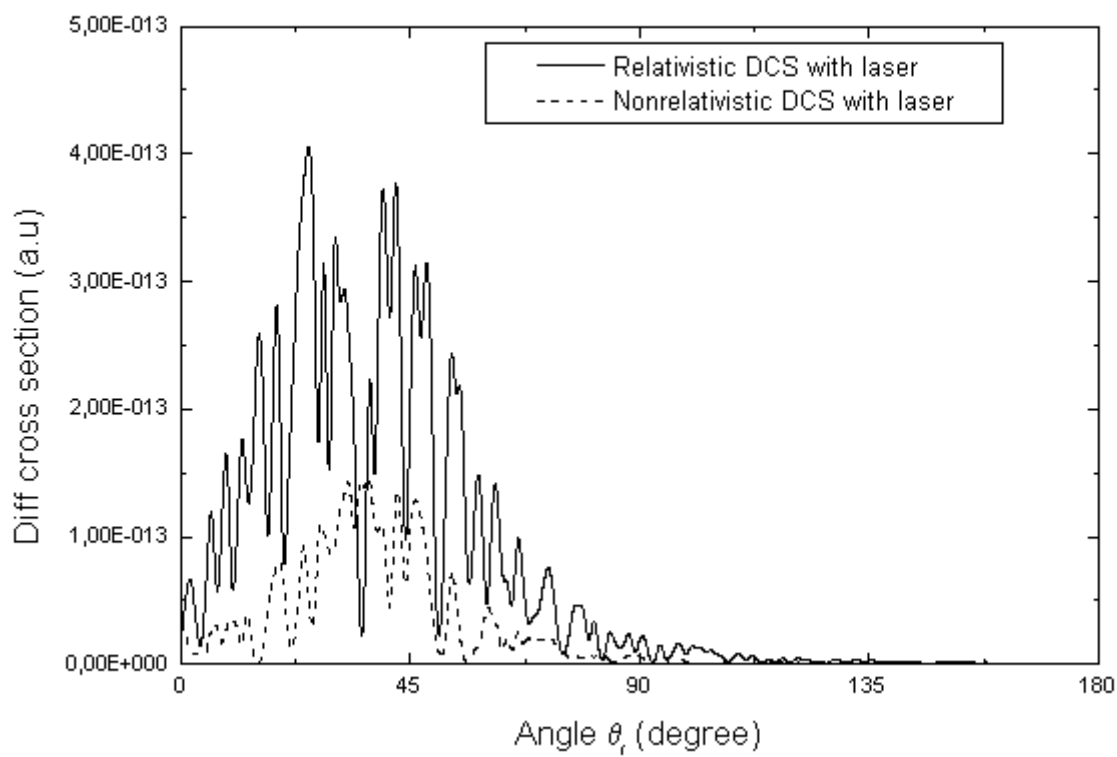}
\caption{Comparison in the relativistic regime ($\gamma=2\,a.u$,
$\varepsilon=1\,a.u$ and $w=0.043\,a.u$) between the relativistic
DCS with laser and the nonrelativistic DCS with laser for
absorption of one photon. The geometry chosen is
$\theta_i=\phi_i=45^{\circ}$ where $\theta_f$ varies from
$0^{\circ}$ to $180^{\circ}$ with $\phi_f=90^{\circ}$}
\end{center}
\end{figure}
\newpage
\begin{figure}

\begin{center}
\includegraphics[angle=0,width=16cm]{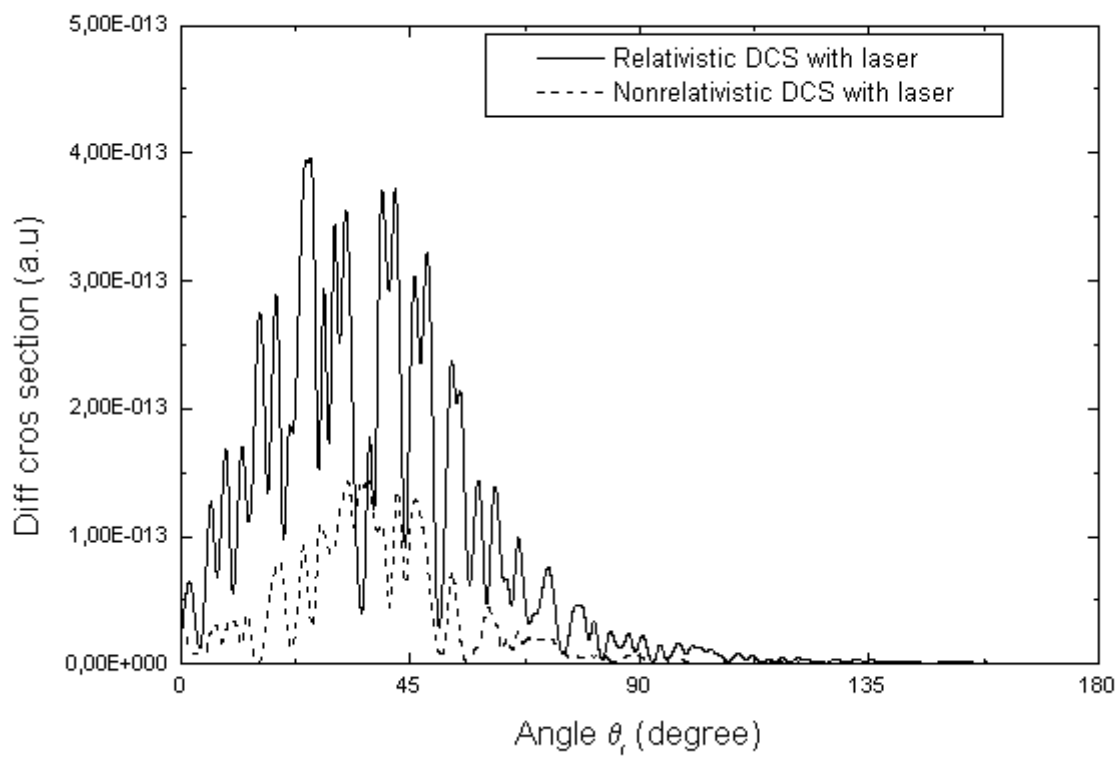}
\caption{Comparison in the relativistic regime ($\gamma=2\,a.u$,
$\varepsilon=1\,a.u$ and $w=0.043\,a.u$) between the relativistic
DCS with laser and the nonrelativistic DCS with laser for
emission of one photon. The geometry chosen is
$\theta_i=\phi_i=45^{\circ}$ where $\theta_f$ held of $0^{\circ}$
to $180^{\circ}$ with $\phi_f=90^{\circ}$. The curves for
absorption and emission are identical.}
\end{center}
\end{figure}
\newpage
\begin{figure}
\begin{center}
\includegraphics[angle=0,width=16cm]{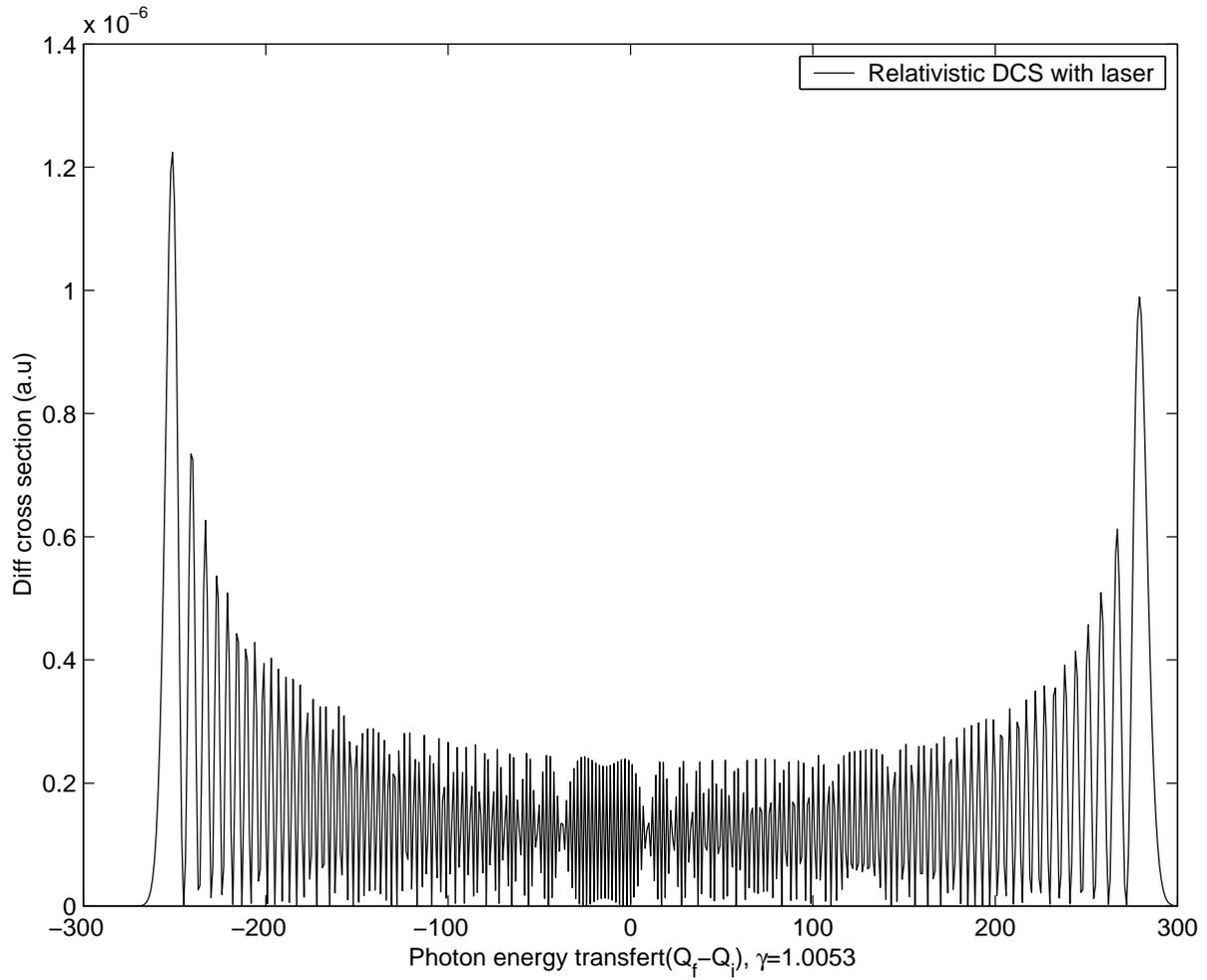}
\caption{The envelope in the nonrelativistic regime
($\gamma=1.0053\,a.u$, $\varepsilon=0.05\,a.u$ and $w=0.043\,a.u$)
of the relativistic DCS with laser for an exchange of $\pm 300$
photons. The geometry chosen is $\theta_i=\phi_i=45^{\circ}$ where
$\theta_f=90^{\circ}$
 with $\phi_f=90^{\circ}$}
\end{center}
\end{figure}
\newpage

\begin{figure}
\begin{center}
\includegraphics[angle=0,width=15cm]{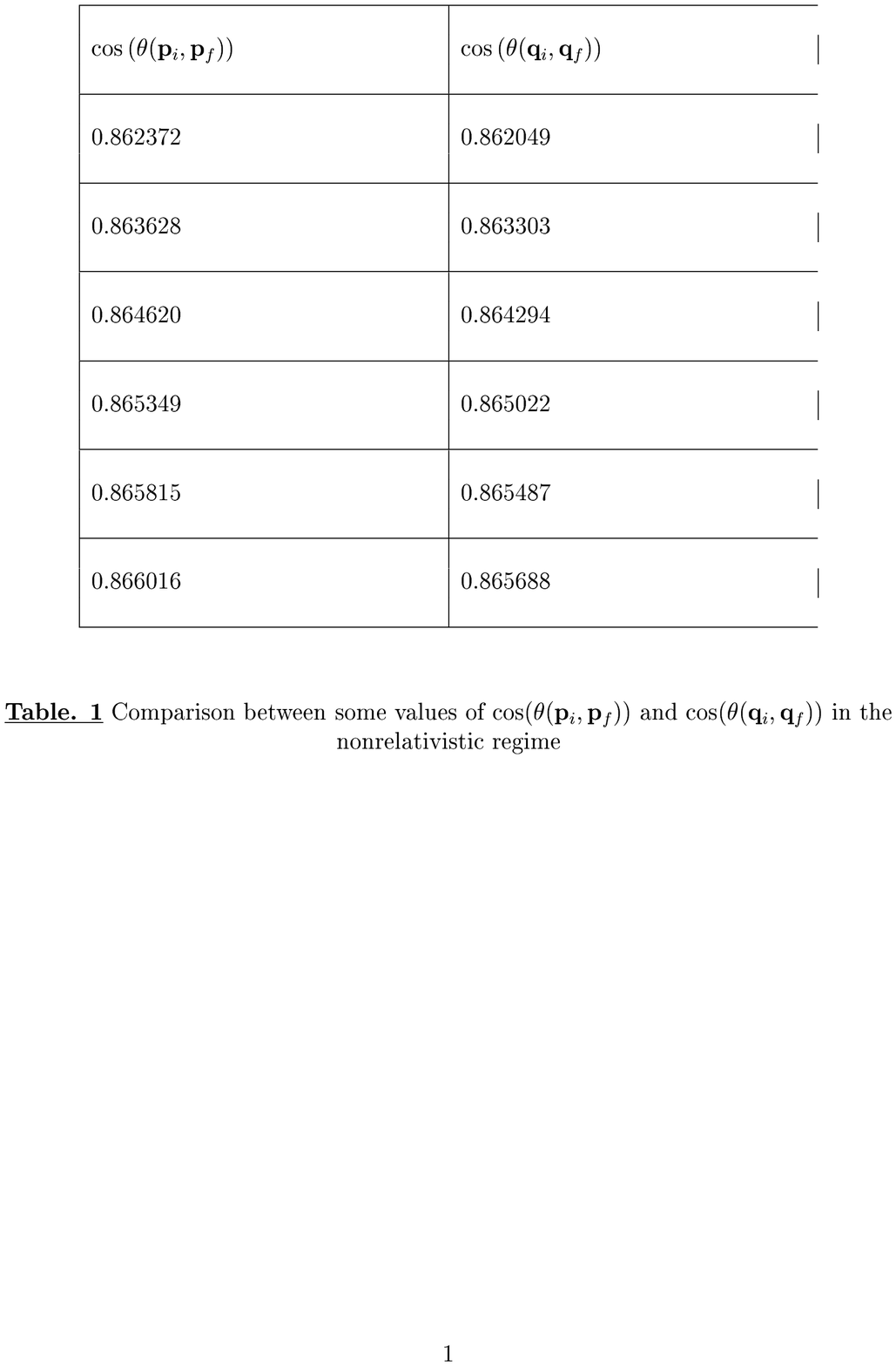}
\end{center}
\end{figure}
\newpage

\begin{figure}
\begin{center}
\includegraphics[angle=0,width=15cm]{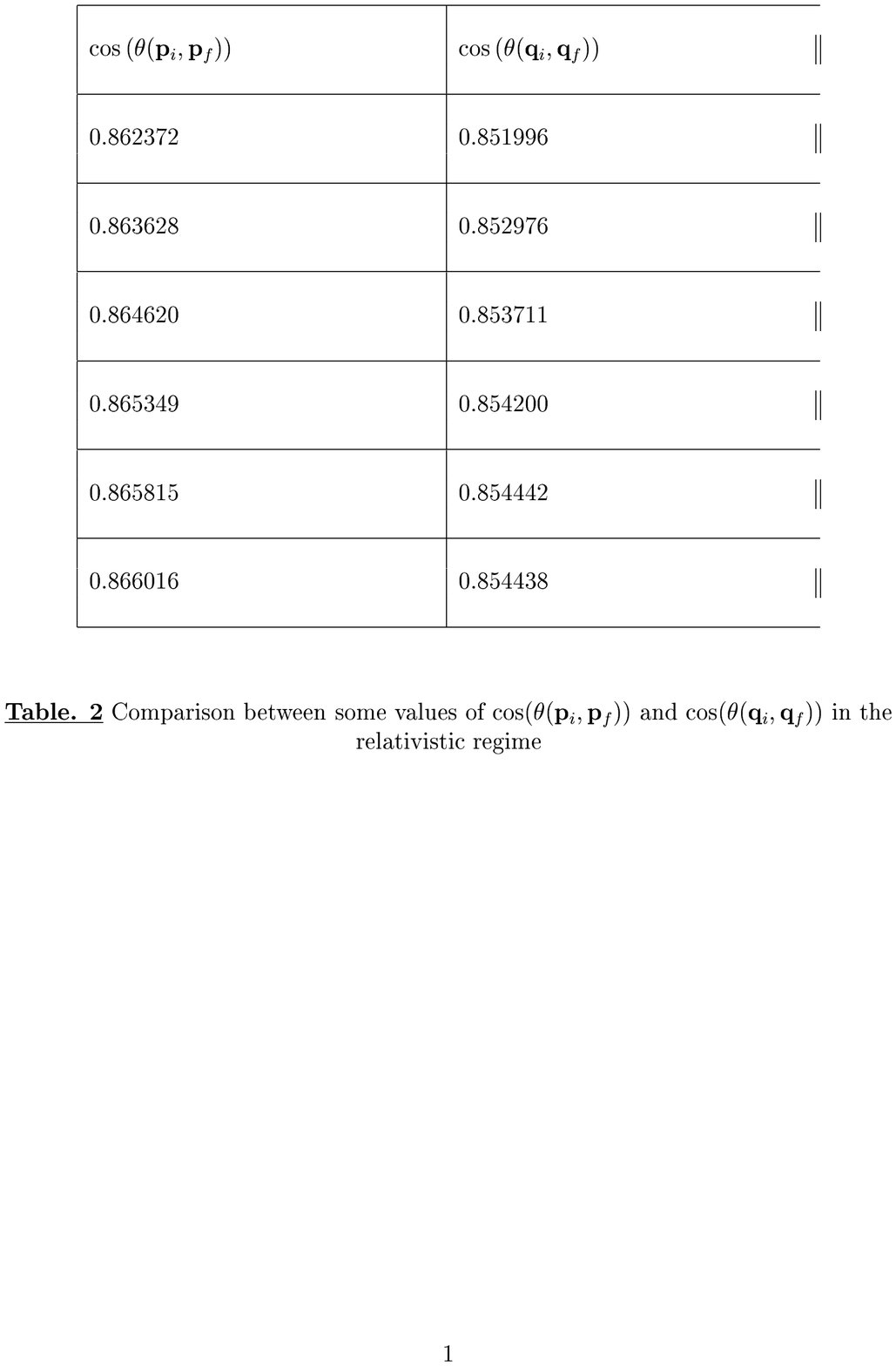}
\end{center}
\end{figure}


\begin{thebibliography}{99}
\bibitem{1}C.I. Moore, J.P. Knauer, and D.D. Meyerhofer,
phys.Rev.lett., \textbf{ 74}, 2953, (1995).
\bibitem{2}C. Bula et al, Phys. Rev.lett. \textbf{76}, 3116
(1996).
\bibitem{3}Y.L. Shao et al, Nature London) \textbf{386}, 54,
(1997).
\bibitem{4}C.J. Joachaine, {\it Quantum collision Theory}, 3rd ed.
(Elsevier Science,1983).
\bibitem{5}F.V. Bunkin and M.V. Fedrov, Soviet. Phys. JETP
\textbf{22}, 844 (1966), N.M. Kroll and K.M. Watson, Phys. Rev. A
\textbf{8}, 804 (1973).
\bibitem{6}F.V. Harteman and A.K. Kerman, Phys. Rev. Lett
\textbf{76}, 624 (1996); F. V. Harteman and N.C. Luhmann, ibid,
\textbf{74}, 1107, (19995).
\bibitem{7}C. Zsymanowski, V. V\'eniard, R. Ta\"{\i}eb, A. Maquet
and C.H. Keitel, Phys. rev A, \textbf{56}, 3846, 1997.
\bibitem{8}J. D. Bjorken and Drell, {\it Relativistic Quantum Mechanics}
(Mac Graw Hill,New York,\\1964); C. Itzykson and J-B Zuber, {it
Quantum field Theory} (Mac Graw Hill,New York,1985).
\bibitem{9}Y. Attaourti, B. Manaut, M. Chabab, hep-ph/0207200.
\bibitem{10}F. W. Bayron, Jr and C.J. Joachain, J. Phys.
B\textbf{17}, L295, (1984)
\bibitem{11}H. Kr\"{u}ger and M.Schulz, J. Phys. B\textbf{9},
1899, (1976); A. D. Gazazian, J. Phys. B\textbf{9},3197 (1976);
N.K. Rahman and F.H.M. Faysal, J. Phys. B B\textbf{11}, 2003,
(1978); M.J. Coneely and S. Geltman, J. Phys B\textbf{14},4847,
(1981); A. lami and N.K. Rahman, J; Phys, B\textbf{16}, L201
(1984); S. Jetzke, F.H.M. Faysal, R. Hippler and O.H. Lutz, Zeit.
Phys, A\textbf{315}, 271, (1984).
\bibitem{12}P. Francken and C.J. Joachain, Phys. Rev A\textbf{35},
1590, (1987); A. Dubois, A. Maquet et S.Jetzke, Phys. Rev A34,
1888(1986), P. Francken, Y. Attaourti, and C.J. Joachain, Phys.
Rev A \textbf{38}, (1988).
\bibitem{13}H. R. Reiss, J. Opt. Soc. Am. B\textbf{7}, 574 (1990).
\bibitem{14}D.P. Crawford and H. R. Reiss, Phys. Rev.
A\textbf{50}, 1844, 1994.

\end{thebibliography}
\end{document}